\documentclass{PoS}
\usepackage{axodraw}

\newcommand{\be}{\begin{equation}}
\newcommand{\ee}{\end{equation}}
\newcommand{\ba}{\begin{eqnarray}}
\newcommand{\ea}{\end{eqnarray}}

\title{Radiative and semileptonic decays in Chiral Perturbation Theory}

\ShortTitle{Radiative and semileptonic decays in Chiral Perturbation Theory}

\author{\speaker{Johan Bijnens}\\ 
        Department of Theoretical Physics, Lund University,\\
        S\"olvegatan 14A, SE 22362 Lund, Sweden\\
        E-mail: \email{bijnens@thep.lu.se}}

\abstract{I give a short overview of what has been done in radiative and
semileptonic Kaon decays in Chiral Perturbation Theory. This includes
for semileptonic decays the work which has been done to order $p^6$ including
preliminary results of isospin breaking to that order in $K_{\ell3}$.
For the radiative decays I mainly present
some results from recent work concerning $K_{\ell3\gamma}$.}

\FullConference{Kaon International Conference\\
		 May 21-25, 2007\\
		 Laboratori Nazionali di Frascati dell'INFN}

\renewcommand{\logo}
{\raisebox{-5mm}
{\parbox{\textwidth}{\begin{flushright}LU TP 07-21\\
July 2007
\end{flushright}
\includegraphics{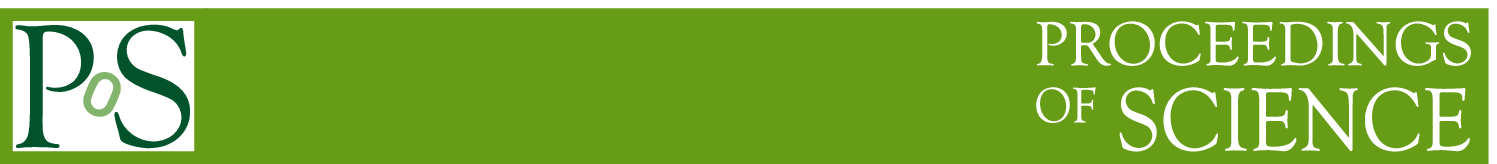}}
}}
\FullConference{{\rm To be published in the proceedings of:\\}
Kaon International Conference\\
		 May 21-25, 2007\\
		 Laboratori Nazionali di Frascati dell'INFN}
\begin{document}

\section{Introduction}

In this talk I give a short overview of the work done on semileptonic
and radiative semileptonic Kaon decays within the context of
Chiral Perturbation Theory (ChPT). There is of course a lot more work done
within ChPT relevant for Kaon decays than I can possibly do justice to
within this short talk. Other theory 
talks with a very large ChPT content presented
at this conference are those of V.~Cirigliano, J.~Prades, J.~Gasser,
G.~Colangelo and C.~Smith.

The Daphne report on semileptonic Kaon decays \cite{daphnereport}
still provides a very good introduction to the ChPT of semileptonic
and radiative semileptonic Kaon decays. Below I will mainly comment
on what has happened since then.

\section{Nonleptonic radiative decays}

\begin{figure}
\vskip-0.5cm
\centerline{\includegraphics[width=7cm]{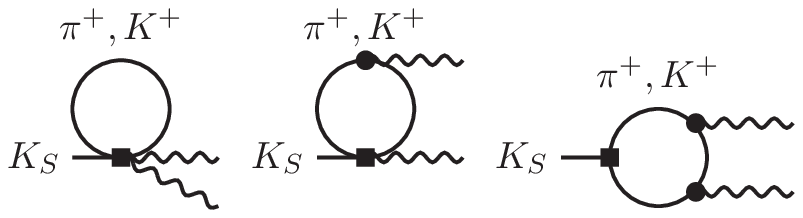}}
\caption{\label{figKSgg}The order $p^4$ diagrams for $K_S\to\gamma\gamma$.}
\end{figure}

Let me first remind you of one of the major successes of ChPT in radiative
decays, the prediction \cite{Goity,DAmbrosio} from the diagrams in
Fig.~\ref{figKSgg} of the decay $K_S\to\gamma\gamma$. The prediction for a
branching ratio of $2.1\cdot 10^{-6}$ has since been amply confirmed by
experiment. There is some disagreement between the NA48 and KLOE
results as discussed in the talk by M.~Martini. However, both results are
in reasonable agreement with the ChPT prediction. There are some estimates of
the effect of final state interactions but no full order $p^6$ calculation
exists.

More experimental results on different nonleptonic radiative
decays can be found in the talks by E.~Imbergamo, E.~Cheu and M.~Martini.
These also discuss comparisons with ChPT.

\section{Semileptonic decays}

Most of these decays were studied in current algebra already.
Gasser and Leutwyler then worked out $K\to\ell\nu$ (really $F_K$)
and $K\to\pi\ell\nu$
in their seminal work on ChPT\cite{GL1,GL2}. The decay $K\to\pi\pi\ell\nu$
was worked out to order $p^4$ in \cite{Riggenbach,Bijnenskl4}
and a dispersive estimate of the higher orders was done by \cite{BCG}.
The latter reference
also evaluated the remaining form-factor $R$ to order $p^4$.
Except for the $R$ formfactor, these calculations have all been
performed to order $p^6$ now and will be discussed in more detail below.
The decay $K_{e5}$ is also kinematically possible. It has been evaluated to
lowest order in ChPT by S.~Blaser \cite{Blaser}. The resulting branching
ratios vary from a few times $10^{-11}$ to a few times $10^{-12}$ and these
processes are thus not very likely to be measured in the foreseeable future.

\subsection{$K_{\ell2}$}

This decay, under the guise of the calculation of $F_K$ and $F_\pi$ for
the corresponding $\pi_{\ell2}$ decays, was performed to order $p^6$
in \cite{ABT1}. Its main use is the fitting of the low-energy constant (LEC)
$L_5^r$ in ChPT. A typical convergence from \cite{ABT4}, fit 10, is
\be
{F_K}/{F_\pi} = 1.22 = 1.000+0.162+0.058\,,\quad\quad{F_\pi}/{F_0}
= 1.000+0.135-0.075\,.
\ee
The 1.22 is used here as input for $F_K/F_\pi$.

\subsection{$K_{\ell3}$}

This decay is one the major sources of knowledge about the parameter $V_{us}$.
This was discussed in great detail at this conference. The theory overview
was given by V.~Cirigliano and a new Flavianet average was presented
by M.~Palutan. The measurements themselves were also discussed by several
other speakers. 

The corrections to $f_+(0)$ at order $p^4$ were first done in \cite{LR} and
to all formfactors in \cite{GL2}. The radiative corrections have been discussed
at this conference by V.~Cirigliano and can be found in \cite{Cirigliano}.
Let me discuss here the calculation in the isospin limit to order $p^6$
\cite{BT} first, some results can also be found in \cite{Post}.
First it was pointed out that $f_+(0)$ is in fact rather
dependent on the curvature in the formfactor $f_+(t)$, as discussed
in the experimental talks here, that curvature has since been measured, thus
removing that uncertainty in the experimental determination of $f_+(0)$
in principle.
Unfortunately, the experiments are not in full agreement due to the large
correlations between the linear slope and the curvature. However, when using
the pole form for $f_+(t)$ all experiments are in good agreement and agree
reasonably well with the ChPT prediction of the curvature\cite{BT}.

One of the problems with ChPT is the large number of parameters that appear at
higher orders. However, for $f_+(0)=f_0(0)$ the needed order $p^6$ LECs,
$C_{12}^r$ and $C_{34}^r$, are in principle determinable from the
scalar formfactor slope and curvature:
\ba
f_0(t) &=& 1-\frac{8}{F_\pi^4}{\left(C_{12}^r+C_{34}^r\right)}
\left(m_K^2-m_\pi^2\right)^2
+8\frac{t}{F_\pi^4}{ \left(2C_{12}^r+C_{34}^r\right)}
\left(m_K^2+m_\pi^2\right)
\nonumber\\&&
+\frac{t}{m_K^2-m_\pi^2}\left(F_K/F_\pi-1\right)
-\frac{8}{F_\pi^4} t^2 { C_{12}^r}
+\overline\Delta(t)+\Delta(0)\,.
\ea
$\overline\Delta(t)$ and
$\Delta(0)$ contain {\em no} $C_i^r$ and only depend on the
$L_i^r$ at order $p^6$.
All needed parameters can thus {\em in principle}
be determined experimentally. In practice this will be difficult,
but dispersion theory can help producing relations between the slope
and curvature,
see e.g. the talk by E.~Passemar at this conference. Thus measuring the
scalar formfactor as precisely as possible should still help in improving
the accuracy for $V_{us}$.
The final conclusion of \cite{BT} for the remainder was
\be
\Delta(0) = -0.0080\pm0.0057[\mbox{loops}]\pm0.0028[L_i^r]\,,
\ee
where there is a sizable cancellation between the order $p^4$ and the
pure loop order $p^6$ contribution. 

As discussed in the overview experimental talk on $K_{\ell3}$
by M.~Palutan, there is a small discrepancy between the determinations
of $V_{us}$ from $K^+_{\ell3}$ and  $K^0_{\ell3}$ decays. The ratio between these
decays is well predicted to order $p^4$ and $e^2p^2$ in ChPT. There is thus a
question whether the remainder is due to isospin breaking at higher order in
ChPT. Such a calculation is underway \cite{KG} and preliminary results
are shown in Tab.~\ref{tab:kl3}. There also the results of the isospin
conserving calculation are shown for $f_+(0)$. What can be seen is that
the order $p^6$ correction in general {\em lowers} the isospin breaking seen in
the ratio. However, when the full order $p^6$ fit including isospin breaking
\cite{ABT4} is taken into account there is an increase by about 0.5\% compared
to the old order $p^4$ results of \cite{GL2,LR}. The change in $m_u/m_d$ comes
about half from the order $p^6$ corrections and half from the inclusion of
corrections to Dashen's theorem.
\begin{table}
\begin{center}
\begin{tabular}{ccccccc}
\hline
Decay & $p^2$ & $p^4$ & pure 2-loop & $L_i^r$ at $p^6$ & $C_i^r$ & total \\
\hline
\multicolumn{6}{c}{Iso conserving calculation}\\
$K^0$ & 1 & $-$0.02266 & 0.01130 & 0.00332 & ??? & 0.99196\\
$K^+$ & 1 & $-$0.02276 & 0.01104 & 0.00320 & ??? & 0.99154\\
\hline
\multicolumn{6}{c}{$m_u/m_d = 0.45$}\\
$K^0$ & 1 & $-$0.02310 & 0.01131 & 0.00325 & ??? & 0.99146\\
$K^+$ & 1.02465 & $-$0.01741 & 0.00379 & 0.00648 & ??? & 1.01751\\
ratio & 1.02465 & 1.0311 &  & & & 1.0262\\
\hline
\multicolumn{6}{c}{$m_u/m_d = 0.58$}\\
$K^0$ & 1 & $-$0.02299 & 0.01124 & 0.00325 & ??? & 0.99150\\
$K^+$ & 1.01702 & $-$0.01897 & 0.00657 & 0.00551 & ??? & 1.01013\\
ratio & 1.0170 & 1.0215 &  & & & 1.0188\\
\end{tabular}
\end{center}
\caption{\label{tab:kl3} {\em Preliminary} results of isospin breaking to
order $p^6$ for $f_+(0)$ in $K_{\ell3}$\cite{KG}.
$m_u/m_d=0.58$ corresponds to the values used in \cite{LR,GL2}
while $m_u/m_d=0.45$ is the result of the full order $p^6$ fit\cite{ABT4}.}
\end{table}
Dealing with $\pi^0$-$\eta$ mixing is somewhat more involved at
order $p^6$ but can be dealt with\cite{ABT4,KG}. Whether all necessary order
$p^6$ LECs can be determined similar to the isospin conserving case is
under investigation\cite{KG}.

\subsection{$K_{\ell4}$}

This decay plays a major role in the discussion of $\pi\pi$ scattering.
This was discussed at this conference by G.~Colangelo,
B.~Bloch-Devaux and L.~Di~Lella, here I concentrate
on the values of the formfactors. The lowest order result,
$
F=G= {m_K}/({\sqrt{2}F_\pi})$,
was worked out by Weinberg.
The order $p^4$ calculations were done around 1990
 \cite{Riggenbach,Bijnenskl4}
and a dispersive estimate of the higher orders appeared somewhat later \cite{BCG}.
The latter reference
also evaluated the remaining form-factor $R$ to order $p^4$.
The absolute values of $F$ and $G$ provide the best input for $L_i^r$,
$i=1,2,3$. The full order $p^6$ calculation was done
in \cite{ABT2,ABT3} for $F$ and $G$ and in \cite{ABBC} for the vector
formfactor $H$. It was found that the full calculation gave a higher order
correction somewhat larger than the dispersive estimate but that the
ChPT series was generally converging. 
An updated fit of the ChPT LECs
to the new NA48 $K_{\ell4}$ measurements and including the
information from $\pi\pi$ scattering is certainly needed.

\section{Radiative semileptonic Kaon decays}

\subsection{$K_{\ell2\gamma}$}

The internal Bremsstrahlung contribution to this decay was
known for a long time. The order $p^4$ contribution was done
in \cite{BCE} and the order $p^6$ in  \cite{Geng} for the axial formfactor.
The vector formfactor is predicted
at order $p^4$ by the anomaly and is known to order $p^6$ \cite{ABBC}.
The corrections are of the expected size for both formfactors.

\subsection{$K_{\ell3\gamma}$}

The inner Bremsstrahlung contribution was calculated using current
algebra methods\cite{FFS}. The order $p^4$ calculation was performed in
\cite{BCE} and a thorough analysis of the higher order corrections
using the fact that cuts are far away from the physical region can be found
in the papers \cite{Kubis1,Kubis3,Kubis4}.
It was found there that in the ratios
\newcommand{\rr}{{R}}
\newcommand{\MeV}{{\,{\rm MeV}}}
\def\ke3g{K_{e3\gamma}}
\newcommand{\Eg}{E_\gamma^*}
\newcommand{\Ecut}{E_\gamma^{\rm cut}}
\newcommand{\Ep}{E_\pi^*}
\newcommand{\Ee}{{E_e^*}}
\newcommand{\te}{\theta_{e\gamma}^*}
\newcommand{\tep}{\theta_{e\pi}^*}
\newcommand{\tng}{\theta_{\nu\gamma}^*}
\newcommand{\tecut}{\theta_{e\gamma}^{\rm cut}}
\be
R\left(\Ecut,\,\tecut\right) =
{\Gamma\left(\ke3g^\pm,\, \Eg > \Ecut,\, \te>\tecut \right)}
/{\Gamma\left(K_{e3}^\pm\right)} ~,  
\ee
many of the uncertainties drop out. Studies of how the structure functions
can be determined experimentally can also be found there. As an example I show
the table for the ratios as a functions of the $K_{\ell3}$ slopes\cite{Kubis1}.
\begin{table}
\begin{center}
{
\renewcommand{\arraystretch}{1.4}
\begin{tabular}{ccccccc}
\hline
$\Ecut$ & $\tecut$ &
$\rr^{\rm IB} \cdot 10^2$ & 
$\rr \cdot 10^2$ & 
$c_1 \cdot 10^3$ & 
$c_2 \cdot 10^4$ & 
$c_3 \cdot 10^4$ \\
\hline 
30\,MeV & $20^\circ$         & 0.640 
                            & $0.633\pm 0.002$ & $12.5\pm 0.4$ & $-5.4\pm 0.3$ & $16.9\pm 0.4$ \\
30\,MeV & $10^\circ$         & 0.925 
                            & $0.918\pm 0.002$ & $11.1\pm 0.3$ & $-4.7\pm 0.2$ & $15.0\pm 0.3$ \\
10\,MeV & $20^\circ$         & 1.211 
                            & $1.204\pm 0.002$ & $7.5\pm 0.2$ & $-3.2\pm 0.2$ &  $10.1\pm 0.2$ \\
10\,MeV & $10^\circ$         & 1.792 
                            & $1.785\pm 0.002$ & $6.7\pm 0.2$ & $-2.8\pm 0.1$ & $9.0\pm 0.1$ \\
10\,MeV & $26^\circ-53^\circ$ & 0.554 
                            & $0.553\pm 0.001$ & $5.7\pm 0.1$ & $-2.4\pm 0.1$ & $7.5\pm 0.1$ \\
\hline 
\end{tabular}}
\end{center}
\caption{\label{tab:kl3g} The ratios $R$ in $K_{\ell3\gamma}$ decay as
a function of the $K_{\ell3}$ formfactor parameters\cite{Kubis1}.}
\end{table}

\section{Conclusions}

Chiral Perturbation Theory has been extremely useful in
semileptonic and radiative Kaon decays. I have given a very fast overview of
existing calculations and presented some preliminary results for the
order $p^6$ isospin breaking in $K_{\ell3}$\cite{KG}.
Let me finish by pointing out that in addition to all the results discussed
above, many of these decays provide good tests of the anomaly, including
its sign \cite{daphnereport,ABBC,Bijnensanomalies}.
Much of what I discussed here has been reviewed to order $p^4$ in
\cite{daphnereport} and to order $p^6$ in \cite{Bijnensreview}.

\acknowledgments

I like to thank the organizers for a very enjoyable meeting and in getting
together a great program in Kaon physics.
This work is supported in part by the European Commission RTN network,
Contract MRTN-CT-2006-035482  (FLAVIAnet), 
the European Community-Research Infrastructure
Activity Contract RII3-CT-2004-506078 (HadronPhysics) and
the Swedish Research Council.

\end{document}